\documentclass[prl,twocolumn,aps,showpacs,superscriptaddress]{revtex4}
\usepackage{graphicx}
\usepackage{dcolumn}
\usepackage{bm}
\usepackage{amssymb}
\usepackage{epsfig}
\begin{document}
\title{Josephson current in nanofabricated V/Cu/V mesoscopic junctions}
\author{C\'{e}sar Pascual Garc\'{i}a}
\email{c.pascual@sns.it}
\affiliation{NEST CNR-INFM and Scuola Normale Superiore, Piazza dei
Cavalieri 7, I-56126 Pisa, Italy}
\author{Francesco Giazotto}
\email{f.giazotto@sns.it}
\affiliation{NEST CNR-INFM and Scuola Normale Superiore, Piazza dei
Cavalieri 7, I-56126 Pisa, Italy}
\date{\today}
\begin{abstract}
We report the fabrication of planar V/Cu/V mesoscopic Josephson weak-links of different size, and the analysis
of their low-temperature behavior. The shorter junctions exhibit critical currents of several tens of $\mu$A at 350 mK,  while Josephson coupling persists up to $\sim 2.7$ K.
Good agreement is obtained by comparing the measured switching currents to a model which holds in the diffusive regime.
Our results demonstrate that V is an excellent candidate for the implementation of superconducting nanodevices operating at a few Kelvin.
\end{abstract}

\pacs{74.78.Na, 74.50.+r, 74.45.+c}

\maketitle
\par
Josephson coupling in superconductor-normal metal-superconductor (SNS) proximity junctions is one of the key manifestations of quantum coherence in mesoscopic systems~\cite{Nregion,nanow,sueur,cortois}.
For this reason it has been attracting a lot of interest  from  both the fundamental and the applied physics point of view~\cite{book}.
Most of \emph{metallic} SNS weak-links are realized within the diffusive regime, and belong typically to the \textit{long}-junction limit which holds for $\Delta \gg E_{Th}$. Here $\Delta$ is the superconducting order parameter, $E_{Th}=\hbar D/L^2$ is the Thouless energy, $D$ is the diffusion coefficient of the N region, and $L$ its length.
In such a limit, $E_{Th}$ sets the zero-temperature critical current of the junction \cite{dubos}, as well as it determines the scale of supercurrent  decay with temperature ($T$) for $k_BT\gg E_{Th}$ \cite{dubos,zaikin}, where $k_B$ is the Boltzmann constant.
Yet, when the temperature is such that $\Delta(T)$ becomes comparable to $E_{Th}$, both these energies turn out to be relevant to control the junction response.
Therefore,  larger Thouless energy as well as enhanced $\Delta$ are required to extend the operation of the SNS junction at higher temperatures.
While the former issue can be solved by shortening $L$, the latter requires the exploitation of S materials with higher critical temperature ($T_c$).
So far, most of nanofabricated planar SNS junctions make a widespread use of aluminum \cite{nanow,sueur,metals,hoss,cortois,crosser,savin}, which limits operation below $T\sim 1.2$ K, or niobium ($T_c\simeq 9.2$ K) \cite{Nregion,dubos,metals,hoss,morpurgo} that,  however, requires more complex fabrication protocols owing to its high melting point~\cite{Nregion,metals,dubos2}.
On the other hand, among elemental superconductors, vanadium (V) has a number of attractive features.
It is a group-V transition metal like Nb and Ta, and its bulk $T_c \simeq 5.4$ K \cite{keesom} allows applications around liquid $^4$He temperatures.
Furthermore, its melting point is low enough to allow evaporation with ease.
These characteristics make in principle V suitable for the implementation of superconducting circuits operating at temperatures accessible with the technology of $^4$He cryostats.

\begin{figure}[t!]
\includegraphics[width=\columnwidth]{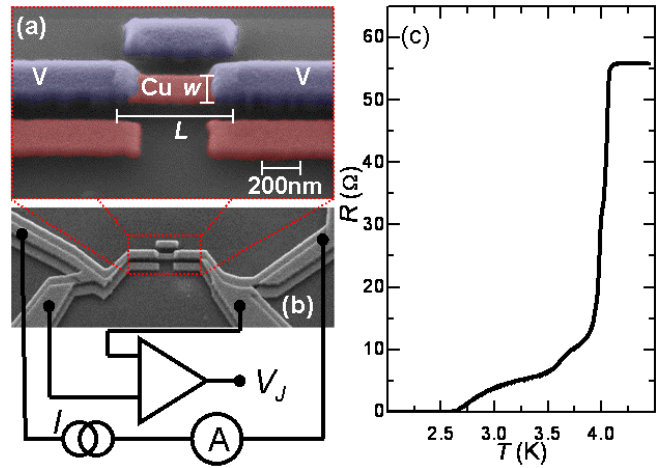}
\caption{\label{fig1} (Color online) (a) Scanning electron micrograph showing the blow-up of a V/Cu/V weak-link in a typical device (sample B). (b) Sketch of the measurement setup. (c) Zero-bias resistance vs temperature for sample A.}
\end{figure}

\begin{table}[b]
\caption{Samples parameters. The junctions total length ($L$), width ($w$), and normal-state resistance ($R_N$) are shown. The experimental Thouless energy ($E_{Th}^{Exp}$) is determined from $R_N$ and the samples geometrical dimensions \cite{param}. The superconducting gap is $\Delta \simeq 0.62$ meV for all samples.
    $E_{Th}^{Fit}$ and the suppression coefficient $\alpha$ are fit parameters of Fig. \ref{fig3}.
    (see text).}
\begin{center}
    \begin{tabular}{ | c  c  c  c  c  c  c  c |}
		\hline
		Sample  & $L$  & $w$ &   $R_N$ & $E_{Th}^{Exp}$ & $\Delta/E_{Th}^{Exp}$ & $E_{Th}^{Fit}$  & $\alpha$ \\
		  &  (nm) & (nm) &   ($\Omega$) & ($\mu$eV) &  & ($\mu$eV)  &  \\
		\hline

			A     & 420 & 184 & 1.55  & 34.3 & 18.1&36.0 &	0.38  \\
			
 			B     & 636 & 189 & 1.62 & 21.1 &29.5& 20.3  & 0.36		\\
 			
 			C     & 783 & 172 & 2.86 & 10.6 &58.4& 10.4  & 0.34		\\
 			
 			D     & 430 &  46 & 7.80  & 26.6 &23.4& 32.7  & 0.125		\\

    \hline
    \end{tabular}
    \label{table1}
\end{center}
\end{table}
In this Letter we report the fabrication of planar V/Cu/V mesoscopic Josephson weak-links, and their characterization down to $T\sim350$ mK.
The samples exhibit sizeable critical currents whose behavior follows what expected in the long-junction limit.
The ease of fabrication combined with the quality of the structures make the V/Cu material system relevant for the realization of superconducting nanodevices.
\par
Samples were fabricated by electron beam lithography and two-angle shadow-mask evaporation onto an oxidized Si substrate.
A 5-nm-thick Ti layer was first evaporated to improve the adhesion of the metallic films to the substrate.
The N region of the weak-link consists of a 40-nm-thick Cu island whose
total length $L$ and width ($w$), depending on the sample, is within the range  $\sim 420\ldots 780$ nm, and $\sim 45 \ldots 190$ nm, respectively.
From the normal-state resistance of the junctions ($R_N$) we extracted the Cu diffusion coefficient which lies in the range $D\sim 74\ldots 130$ cm$^2$/s for the four samples.
Vanadium was evaporated as the last step at a high deposition rate (10 \AA/s ) to avoid oxygen contamination which can lower its critical temperature \cite{vanadium}.
The thickness of the  V film was 120 nm for all samples, and was optimized in order to maximize $T_c$. The main parameters of the four samples are summarized in Table~\ref{table1}.
The junctions were characterized down to $T\simeq 330$ mK in a filtered $^3$He cryostat.
\begin{figure}[t]
\includegraphics[width=\columnwidth]{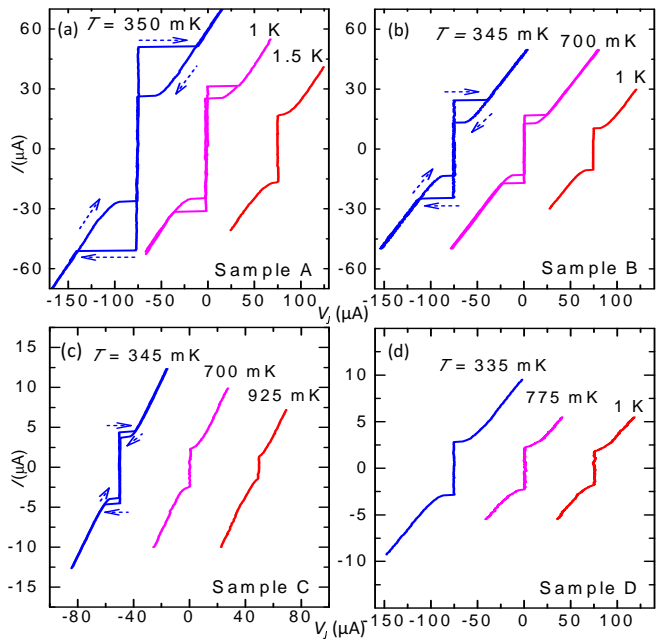}
\caption{\label{fig2} (Color online) Current vs voltage characteristics of all samples at three different representative temperatures. The curves are horizontally offset for clarity.}
\end{figure}

Figure~\ref{fig1} (a) displays the blow-up of a typical V/Cu/V weak-link  (sample B).
The V leads overlap the Cu island laterally for $\sim 100$ nm in all the samples.
A micrograph of the same device along with a scheme of the measurement setup used for the  characterization is shown in Fig.~\ref{fig1}(b).
It can be noted the presence of  sample portions consisting of Ti/V and Ti/Cu/V layers nearby the junction.
Figure~\ref{fig1}(c) shows the zero-bias differential resistance ($R$) vs  bath temperature for sample A.
Above $T$ = 4.2 K the resistance corresponds to that of the whole device in the normal state, while the first transition occurring at $T\simeq 4.1$ K sets the critical temperature of the V film. 
From the latter we deduced a superconducting energy gap $\Delta =1.764 k_B T_c\simeq 0.62$ meV \cite{noer}.
The double transition, occurring around 3.7 K and 3 K, can be ascribed  to the critical temperatures of the metallic bilayer and triple-layer present in the structures [see Fig.~\ref{fig1}(b)].
Then, below $T\simeq 2.75$ K, the zero-bias resistance drops to zero indicating that the junction is in the Josephson regime, and an observable supercurrent is established.
The four samples exhibited a very similar behavior, apart from the transition temperature  to the supercurrent state.

Figure~\ref{fig2} shows the current-voltage ($I$ vs $V_J$) characteristics of all the samples  at three different representative temperatures.
The curves are horizontally offset for clarity.
In particular, for sample A at $T$ = 350 mK [see panel (a)] the $I$ vs $V_J$ displays a clear Josephson effect with a switching current $I_s$ = 52 $\mu$A.
It also shows a marked hysterical behavior, i.e., once the weak-link has switched to the resistive state it recovers the dissipationless regime at  a much smaller bias current, i.e., the retrapping current ($I_r$).
In sample A, $I_r$ = 27 $\mu$A at $T$ = 350 mK.
The origin of hysteresis in mesoscopic Josephson junctions with large critical current has been recently investigated by Courtois \textit{et al.}~\cite{cortois}, and stems from electron heating in the N region once the junction switches to the dissipative branch.
Furthermore, the hysteresis decreases at higher bath temperature, and disappears around $T$ = 1.5 K.
Notably, sample A obtains $I_s=17\,\mu$A at $T=1.5$ K.
A similar general behavior is  also observed  for the current-voltage characteristics of sample B, shown in Fig.~\ref{fig2}(b). The switching current is lower in this case due the increased length of the N wire, and  also its hysterical behavior is less pronounced.
Following the same trend, the current-voltage characteristics of sample C [see Fig.~\ref{fig2}(c)] show a lower $I_s$, and a much reduced hysteresis  has been observed only at the lowest temperatures.
In sample D, the N-region length  is similar to that of sample A but its width is narrower which results in a higher $R_N$. The switching current is therefore lower, and the hysteresis is nearly absent at all the investigated temperatures [see Fig.~\ref{fig2}(d)].
\par
The impact of the N-region size on the supercurrent is displayed in Fig.~\ref{fig3} which summarizes the results obtained in all four devices.
Full dots and open squares represent the $I_s$'s and $I_r$', respectively, extracted from the junctions current-voltage characteristics taken at different temperatures.
The decay of $I_s$ and the softening of the hysteresis with $T$ is  evident.
As explained above, the switching current and the hysteresis are larger for sample A [Fig.~\ref{fig3}(a)], and decrease as $R_N$ increases.
As expected \cite{dubos,zaikin}, the decay of $I_s$ with temperature is also faster for longer junctions (sample B and C), i.e., those with a lower $E_{Th}$ [see Fig. 3(b) and (c)].
Meanwhile sample D shows almost no hysteresis,
and its supercurrent is less affected by the temperature than samples B and C as expected from its higher $E_{Th}$ [Fig. ~\ref{fig3}(d)].
Furthermore, $I_r$ saturates below a threshold temperature which is higher for junctions with larger $I_s$.
We remark the persistence of the Josephson coupling in all samples up to high $T$ (in sample A, for instance, $I_s$ is clearly observable up to $T\simeq 2.7$ K).
This points up the advantage of using V as superconductor to extend the junctions operation at higher temperatures.
\par
\begin{figure}
\includegraphics[width=\columnwidth]{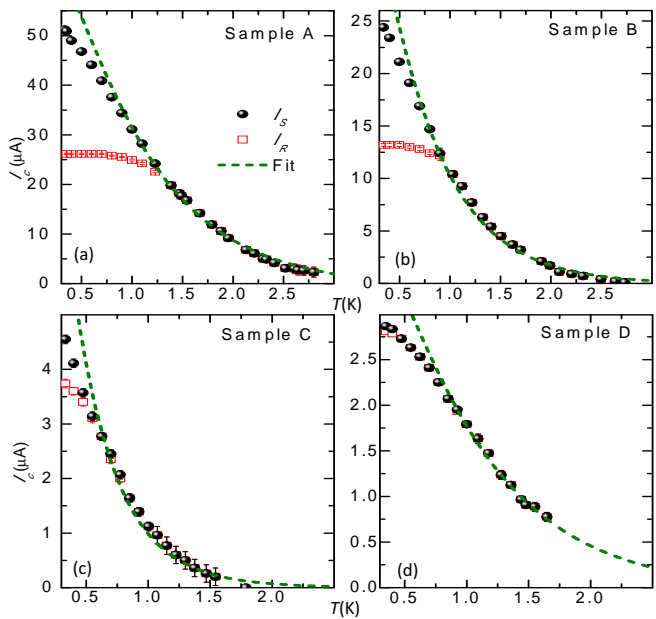}
\caption{\label{fig3} (Color online) Critical current $I_c$ vs $T$ for all four devices. Full dots and dashed lines represent the  switching current ($I_s$) and the high-temperature fit, respectively. The retrapping current ($I_r$) is also shown as open squares.}
\label{fig3}
\end{figure}
The examination of Table~\ref{table1} shows that $\Delta/E_{Th}^{Exp}\gg 1$ which provides the frame of the \textit{long}-junction limit. Here $E_{Th}^{Exp}$ is the experimental Thouless energy which has been
determined from the junctions geometrical dimensions and $R_N$ \cite{param}.
In order to make a more quantitative analysis of the  data we recall that the critical current of a long diffusive  Josephson junction  in the high-temperature regime ($k_BT\gtrsim5E_{Th}$) is given by \cite{zaikin,dubos}
\begin{equation}
I_s(T)=\frac{64\pi k_B T}{eR_N}\sum_{n=0}^{\infty}\frac{\sqrt{\frac{2\omega _n}{E_{Th}}}\Delta ^2(T)\text{exp}[-\sqrt{\frac{2\omega _n}{E_{Th}}}]}{[\omega_n+\Omega_n+\sqrt{2(\Omega_n^2+\omega_n\Omega_n)}]^2},
\label{eqsup}
\end{equation}
where $\omega_n(T)=(2n+1)\pi k_BT$, and $\Omega_n(T)=\sqrt{\Delta^2(T)+\omega_n^2(T)}$.
Dashed lines in Fig.~\ref{fig3} represent the best fit to the data calculated from Eq. (\ref{eqsup}) by setting the nominal $R_N$ and $\Delta$ values, and using as
fitting parameters the Thouless energy ($E_{Th}^{Fit}$)  and a suppression coefficient ($\alpha$) \cite{cortois} multiplying Eq. (1) which accounts for nonideality of the junction.
For $\Delta(T)$ we used the usual BCS temperature dependence of the gap.
For all devices good agreement with the data is obtained by setting $E_{Th}^{Fit}$ very close to the experimental estimate with $\alpha \sim 0.35$ for the first three samples (A-C), and $\alpha =0.125$ for sample D (see Table~\ref{table1}).
The origin of $I_s$ suppression \cite{cortois,metals,savin} is at present not fully understood, but
such values for $\alpha$ \cite{cortois} may suggest the presence of enhanced scattering due to disorder in the junction, or at the SN interface which could be intrinsic to the nature of V/Cu metallurgical  contact.
As suggested by M. Yu. Kupriyanov \cite{kup} a plausible reason for this suppression stems from the overlapped geometry typical of shadow-mask evaporated structures, which leads to a more complex system than the SNS one described by Eq. (\ref{eqsup}), in particular at both NS boundaries which consist of layered proximized regions.  
We note finally that this deviation from theoretical prediction of Eq. (\ref{eqsup}) is not to be ascribed to quasiparticle overheating in the N region, due to the strong electron-acoustic phonon coupling in such a high temperature regime \cite{rev}.

\par
In conclusion, we have reported the fabrication of V/Cu/V proximity Josephson junctions, and their characterization at cryogenic temperatures.
The weak-links are easy to fabricate with present-day technology, and show sizeable  critical currents at temperatures of a few Kelvin.
The supercurrents decay with temperature is in good agreement with what expected in the diffusive regime, while their  amplitude shows a reduction which could be related to the nature of the V/Cu contact.
Our results demonstrate that V/Cu/V mesoscopic junctions are an attractive choice for the implementation of a broad range of superconducting devices, for instance, from SQUIDS \cite{sueur,book,metals} and  out-of-equilibrium Josephson transistors \cite{Nregion,crosser,savin,morpurgo,rev} to radiation sensors \cite{PJS}.
\par
We acknowledge T. T. Hekkil\"a, J. P. Pekola, and F. Taddei for fruitful discussions, M. Yu. Kupriyanov for pointing out a possible mechanism for supercurrent suppression in our Josephson junctions, 
and partial financial support from the NanoSciERA "NanoFridge" project.


\begin{thebibliography}{99}
\bibitem{param}
We recall that $D=L/(\nu_Fe^2twR_N)$ where $t=40$ nm is the N island thickness, and $\nu_F=1.56\times 10^{47}$ J$^{-1}$m$^{-3}$ is the density of states at the Fermi level of Cu.
\end{thebibliography}
\end{document}